\pdfoutput=1
\documentclass{article}
\usepackage{locata,amsmath,graphicx,url,times}

\usepackage{amsfonts}
\usepackage{amsthm} % Theorem Formatting
\usepackage{amssymb}	% Math symbols such as \mathbb
\usepackage{eqnarray}
\usepackage{cite}

\newcommand{\be}{\begin{equation}}
\newcommand{\ee}{\end{equation}}

\newcommand{\abs}[1]{\left| #1 \right|} % for absolute value
\let\baraccent=\= % rename builtin command \= to \baraccent
\renewcommand{\=}[1]{\stackrel{#1}{=}} % for putting numbers above =

\theoremstyle{definition}

\theoremstyle{remark}

\usepackage[acronym]{glossaries}
\usepackage{balance}
\usepackage[font=small,labelfont=bf]{caption}

\newacronym{bte}{BTE}{behind-the-ear}
\newacronym{ha}{HA}{hearing aid}
\newacronym{hrtf}{HRTF}{head-related transfer function}
\newacronym{rtf}{RTF}{room transfer function}
\newacronym{imtf}{IMTF}{inter-microphone transfer function}
\newacronym{imtd}{IMTD}{inter-microphone time differences}
\newacronym{imtds}{IMTDs}{inter-microphone time differences}
\newacronym{snr}{SNR}{signal-to-noise ratio}
\newacronym{ff}{FF}{free field}
\newacronym{imp}{IMP}{inter-microphone phase}
\newacronym{impd}{IPD}{inter-microphone phase difference}
\newacronym{sdr}{SDR}{target signal to diffuse ratio}
\newacronym{doa}{DoA}{direction-of-arrival}
\newacronym{tdoa}{TDoA}{time difference of arrival}
\newacronym{cc}{CC}{cross-correlation}
\newacronym{pm}{CIMP}{CIMP}
\newacronym{cimp}{CIMP}{Circular statistics-based Inter-Microphone Phase difference estimation} 
\newacronym{cimpl}{CIMPL}{Circular statistics-based Inter-Microphone Phase difference estimation Localizer} 
\newacronym{gcc}{GCC}{generalized cross correlation}

\title{Circular statistics-based low complexity DOA\\ estimation for hearing aid application}
\name{L. D. Mosgaard, D.~Pelegrin-Garcia, T. B. Elmedyb, M. J. Pihl, P. Mowlaee}
\address{Widex A/S, Nym\o llevej 6, DK-3540 Lynge, Denmark\\ 
\it {lmos@widex.com}}

\begin{document}
\ninept
\maketitle
\begin{abstract}
The proposed \gls{cimpl} method is tailored toward binaural hearing aid systems with microphone arrays in each unit. The method utilizes the circular statistics (circular mean and circular variance) of \gls{impd} across different microphone pairs. These \gls{impd}s are firstly mapped to time delays through a variance-weighted linear fit, then mapped to azimuth \gls{doa} and lastly information of different microphone pairs is combined. The variance is carried through the different transformations and acts as a reliability index of the estimated angle. Both the resulting angle and variance are fed into a wrapped Kalman filter, which provides a smoothed estimate of the \gls{doa}. The proposed method improves the accuracy of the tracked angle of a single moving source compared with the benchmark method provided by the LOCATA challenge, and it runs approximately 75 times faster.

\end{abstract}
\section{Introduction}\label{sec:intro}
% Written by Pejman Mowlaee
% updated: 12.04.2018
% Last updated: 20.07.2018
%\begin{itemize}
% \item Opening paragraph: Importance of multi-microphone speech enhancement as spatial filtering in hands-free speech communication 
% multi-channel phase estimation for anything ... transfer function estimation 
% \item Importance of DOA estimation: 
% \item Difficulties in DOA estimation: errors in DOA eventually limits the achievable performance. Methods for robust adaptive beamforming have been proposed to provide robustness against the errors in the source DOA. 
% \item Conventional Methods for DOA estimation: CC-based methods including GCC~\cite{Knapp1976}, 
% \item The advent of circular statistics 
% \item signal-to-diffuse ratio (SDR) estimation
% \item In this paper paragraph [DONE]
%\item Organization of the paper [DONE]
%\end{itemize}
% \textbf{Opening paragraph:} 
% General introduction
Microphone array processing is of interest for hands-free communication, hearing aids, robotics and immersive audio communication systems. It is used in a wide range of applications including noise reduction~\cite{Schwarz2015,Chakrabarty2018a}, informed spatial filters for source separation~\cite{Thiergart2014,Chakrabarty2018a}, source localization~\cite{Farmani2017} and robust beamforming~\cite{Jarrett2011,SpringerHandbook}. The achievable performance in these applications is heavily governed by the accurate information about the direction-of-arrival (\gls{doa}) of the target source(s).
% DOA estimation overview

Conventional methods for \gls{doa} estimation can be grouped into two classes: i) subspace methods relying on e.g. steered-response power phase transform (SRP-PHAT)~\cite{Dibiase2000}, MUSIC~\cite{Schmidt1986} and ESPRIT~\cite{RoyK89}, and ii) cross-power spectrum phase (CSP) based methods~\cite{Knapp1976,Omologo1996}. While the methods in the two groups are different in terms of their \gls{doa} estimation accuracy and the computational efficiency, among them, CSP is popular due to simplicity and reliability. Of particular importance is the so-called \gls{gcc} method using the phase transform (PHAT) normalization~\cite{Knapp1976} for its robustness in \gls{doa} estimation for acoustic source localization~\cite{Omologo1996}. More recently, circular statistics has shown a great potential in multi-channel source tracking for both subspace-based~\cite{Taseska2017} and CSP-based~\cite{Traa2014} methods.
% \subsection*{In this paper paragraph}

In this paper, we propose CSP-based \gls{doa} estimator which relies on circular statistics throughout all estimation stages (Figure~\ref{fig:cimpl_system_diagram}). Our proposed method, \emph{\gls{cimpl}}, is particularly targeted for application in hearing aids. Specifically, we consider a binaural hearing aid setup consisting of two microphones per hearing aid with a binaural radio connection between each hearing aid. For \gls{doa} estimation in such a hearing aid setup, two major challenges are i) the restricted positioning of microphones with a small microphone inter-spacing on each hearing aid and ii) strict computational limitations. We demonstrate the performance of the proposed method with hearing aid recordings in the presence of a single static source (task 1), a single moving source (task 3) and a single moving source with a moving listener (task 5) as defined in the LOCATA challenge~\cite{LOCATA2018a}.
% We first show the phase estimation bias problem in the CSP methods, particularly for short inter-microphone spacings e.g. hearing aids application (Section 2). The proposed method outputs the mean phase and gives access to the circular moments. Using the circular moments allows us to quantify sound properties such as diffuseness. This is facilitated by introducing the inter-microphone phase difference (IMPD) space (Figure 1). Simulations demonstrate the robustness of the proposed method and its reliability for small microphone inter-spacings where CSP methods have a limited performance\footnote{Here, we assume that the \gls{ims} is known and free- and far-field conditions uphold unless otherwise stated.}.
% \subsection*{Organization of the paper}
% \indent The rest of the paper is organized as follow; In Section~\ref{sec:MotivationBackground} describes some background on the well-known cross-correlation based methods for phase estimation and our motivation to propose the new method relying on the concept of the inter-microphone phase difference (IMPD) space. In Section~\ref{section} we present the proposed method for estimation of the inter-microphone phase difference relying on circular statistics. Section~\ref{sec:results} presents the performance of the proposed method in terms of its robustness to noise and its accuracy in measuring SDNR compared benchmark. obtained by the proposed method and quantifies its robustness versus the cross-correlation based method. Discussions and conclusions are presented in Section \ref{sec:discussions}.

\section{DOA estimation}
\begin{figure}[t]%
\centering
%Trimming figure as visio exports margins (at least in this case), usage: trim={<left> <lower> <right> <upper>}
\includegraphics[trim={1cm 0.8cm 1cm 0cm},clip=true,width=8.5cm]{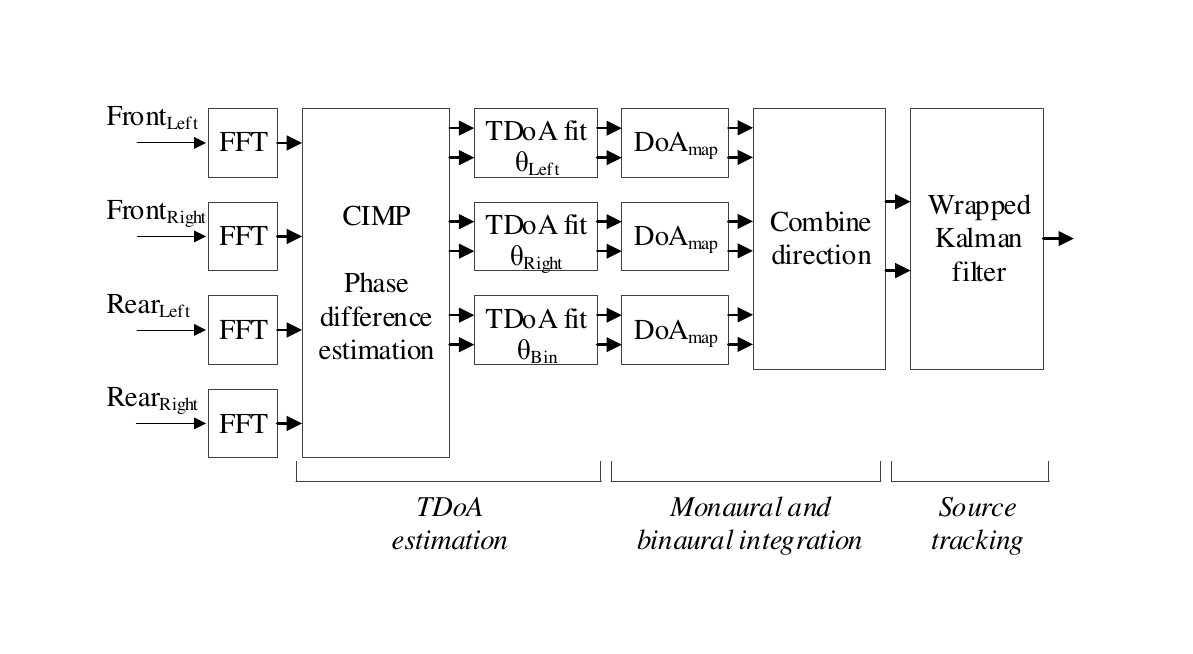}%
\vspace{-0.25cm}
\caption{{System diagram for the proposed method composed of three stages: i) \gls{tdoa} estimation relying on \gls{cimp} and \gls{tdoa} fit to left, right and binaural \gls{impd}s, ii) data association by integrating the monaural (left and right) and binaural \gls{tdoa}s, and iii) source tracker using wrapped Kalman filter.}}
\label{fig:cimpl_system_diagram}%
\end{figure}

%system_overview.tex
The \gls{cimpl} method is based on three major components: i) \gls{tdoa} estimation, ii) monaural and binaural integration, iii) and source tracking. Figure~\ref{fig:cimpl_system_diagram} provides an overview of the \gls{cimpl} method. The different stages are explained in the following. 
\subsection{Time difference of arrival estimation}
The initial step in \gls{cimpl} is to estimate the \gls{tdoa} for each microphone set. The \gls{tdoa} estimation is divided in two stages operating in the frequency domain. The first stage is a phase difference estimation and the second stage consists of a weighted linear fit to estimate the \gls{tdoa}. 

\subsubsection{Circular statistics-based inter-microphone phase difference estimation (CIMP)}\label{sec:CIMP}
%As a circular random variable, the inter-microphone phase (IMP) denoted by $\psi$, its statistical properties is governed by circular statistics. The mean of $\psi$ is given by \cite{Fisher,Mardia}  
%\be
%\mathbb{E} \{ e^{-j \psi} \} = R e^{-j\mu},
%\label{eq:cstat}
%\ee
%with $\mu$ as the circular mean of $\psi$ and $R\in[0,1]$ as the mean resultant length. The latter gives some insight into the spread of $\psi$ and is related to the circular variance ($V = 1 - R$). 
The instantaneous \gls{impd} at frame $l$ and frequency bin $k$, denoted by $\theta_{ab}(k,l)$, defined between two microphones $a$ and $b$ is given by the instantaneous normalized cross-spectrum
\be
e^{j\theta_{ab}(k,l)} = \frac{X_a(k,l) X_b^*(k,l) }{\abs{X_a(k,l)X_b(k,l)}},
\label{eq:instimpd}
\ee
where $X_a$ and $X_b$ are the short-time Fourier transforms of the input signals at the two microphones and $j=\sqrt{-1}$. We assume that $\theta_{ab}(k,l)$ is a particular realization of a circular random variable $\Theta$. Therefore, the statistical properties of the \gls{impd}s are governed by circular statistics and the mean is given by \cite{Fisher,Mardia}
\be
 \mathop{\mathbb{E}}\limits_{l}\{e^{j\theta_{ab}(k,l)}\} = R_{ab}(k,l) e^{j\hat{\theta}_{ab}(k,l)},
\label{eq:mpd}
\ee
where $\mathbb{E}$ is a short-time expectation operator (moving average), $\hat{\theta}_{ab} \in [-\pi,\pi[$ is the mean \gls{impd} and $R_{ab} \in [0,1]$ is the mean resultant length.
% We note here that \eqref{eq:mpd} can be written in the form of the so-called GCC-PHAT function \cite{Knapp1976},
% \begin{eqnarray}
% \tau & = &  \argmax \left\{ \mathbb{E} \left\{F^{-1}\left( e^{-j\theta_{ab}(f)}  \right )\right\} \right\}\nonumber \\
%         & = & \argmax \left\{F^{-1}\left(\mathbb{E}\{e^{-j\theta_{ab}(f)}\} \right) \right\},
%         \label{eq:gcc-phat}
% \end{eqnarray}
% where $F^{-1}$ is the inverse discrete Fourier transform and $\tau$ is the \gls{tdoa} for which the expression is maximized. From this, the GCC-PHAT is re-casted as the appropriate circular statistical method for estimating the \gls{tdoa}, where each frequency is weighted according to the mean resultant length. The GCC-PHAT is known to have excellent performance in noise scenarios where the noise can be considered independent at the microphones but it suffers for small microphone arrays in diffuse noise \cite{Omologo1996,Nesta2008,Loesch2010}.

The mean resultant length carries information about the directional statistics of the impinging signals at the hearing aid, specifically about the spread of the \gls{impd}. For uniformly distributed $\Theta$, which corresponds to the signal at the two microphones being completely uncorrelated, the associated mean resultant length goes to 0. At the other extreme $\Theta$ is distributed as a Dirac delta function $\Theta \sim \mathcal{W}\left\{\delta(\theta_{ab}-\theta_0)\right\}$ corresponding to an ideal anechoic source for a specific frequency $f$ at $\theta_0 = 2 \pi f d/c \cos{\varphi}$, where $\mathcal{W}\left\{\cdot\right\}$ denotes the transformation that maps a probability density function to its wrapped counterpart \cite{Fisher}, $d$ is the inter-microphone spacing, $c$ is the speed of sound, and ${\varphi}$ is the angle of arrival relative to the rotation axis of the microphone pair. In this case, the mean resultant length converges to one. 

A particular detrimental type of interference, both for speech intelligibility and for common \gls{doa} algorithms, is late reverberation typically modeled as diffuse noise. Diffuse noise is characterized by being a sound field with completely random incident sound waves~\cite{Cook1955}. This corresponds to the \gls{impd} having a uniform probability density $\Theta \sim \mathcal{W}\left\{\mathcal{U}(-\pi f / f_u, \pi f / f_u)\right\}$,
%$p_d(\theta) \in [-\pi f / f_u, \pi f / f_u]$}
where $f_u=c/(2d)$ is the upper frequency limit where phase ambiguities, due to the $2\pi$-periodicity of the \gls{impd}, are avoided. For diffuse noise scenarios, the mean resultant length for low frequencies ($f << f_u$) approaches one. It gets close to zero as the frequency approaches the phase ambiguity limit. Thus, at low frequencies, both diffuse noise and localized sources have similar mean resultant length and it becomes difficult to statistically distinguish the two sound fields from each other.
% , which leads to the aforementioned issue of GCC-PHAT at small array \gls{tdoa} estimation (for more details we refer to \cite{Omologo1996,Nesta2008,Loesch2010}).\\ 
%Behavior also known from the classic coherence measure, though mean resultant length and coherence are not equivalent. 

To resolve the aforementioned limitation, we propose transforming the \gls{impd} such that the probability density for diffuse noise is mapped to a uniform distribution $\Theta \sim \mathcal{U}[-\pi,\pi[$ for all frequencies up to $f_u$ while preserving the mean resultant length of localized sources. Under free- and far-field conditions and assuming that the inter-microphone spacing is known, the \emph{mapped} mean resultant length $\tilde{R}_{ab}(k,l)$, which is the mean resultant length of the transformed \gls{impd}, takes the form
\be
\tilde{R}_{ab}(k,l) = \abs{\mathop{\mathbb{E}}\limits_{l}\left\{  e^{j\theta_{ab}(k,l) k_u/k} \right\} },
\label{eq:unwrapR}
\ee
where $k_u = 2K f_u /f_s$ with $f_s$ being the sampling frequency and $K$ the number of frequency bins up to the Nyquist limit.
 
The mapped mean resultant length for diffuse noise approaches  zero for all $k < k_u$ while for anechoic sources it approaches one as intended.

Commonly used methods for estimating diffuse noise (e.g., \cite{Allen1977,Westermann2013}) are only applicable for $k>k_u$. Unlike those methods, the mapped mean resultant length works best for $k<k_u$ and is particularly suitable for arrays with very short microphone spacing such as hearing aids. Particularly, by employing the proposed mapped mean resultant length instead of the mean resultant length, correct weighting is applied in time-frequency which takes into account the diffuse noise for low frequency \gls{tdoa} estimation for small microphone arrays like hearing aid.
% as in e.g. GCC-PHAT (\eqref{eq:gcc-phat}), 

Due to the acoustical nature of  hearing aid arrays, only frequencies up to $k_u$ are considered. At higher frequencies, both for the small spacing between the two microphones on one hearing aid (i.e., monaural case) and across the ears (i.e., binaural case), the assumptions of free- and far-field break down.  
%We see that following the teachings of circular statistics the \gls{impd} estimator (Eq. \eqref{eq:mpd}) for each sample from which the expectation value is estimated is normalized and we see that no straightforward simplifications can be done like seen in Eq. \eqref{eq:gcc}.
%\be
%e^{-i\hat{\theta}_{GCC-PHAT}} = \frac{\mathbb{E}\left(M_i(f) M_j^*(f) \right)}{\abs{\mathbb{E}\left(M_i(f) M_j^*(f) \right)}}.
%\label{GCC-PHAT}
%\ee

\subsubsection{Estimating time difference in the frequency domain}
Given the mean \gls{impd} and the mapped mean resultant lengths calculated so far, the \gls{tdoa} corresponding to the direct path from a given source needs to be estimated. In free- and far-field conditions the \gls{tdoa} of a single stationary broadband source corresponds to a constant group delay across frequency, which reduces the problem of estimating the \gls{tdoa} to fitting a straight line $\theta(f) = 2 \pi f \tau$. This is effectively done in \gls{gcc} method by using the inverse Fourier transform and finding the \gls{tdoa} as the time lag that maximizes the \gls{gcc}. %However, performing the fit directly in the frequency domain offers a number of advantages, one being that it is relatively straight forward to take into account knowledge of the acoustic scenario and the signal statistics.
%, examples of utilizing this approach is (cite ransac?). 
%However, in realistic sound scenarios not all these criteria can be met for all frequencies simultaneously and optimally only time-frequency bins dominated by the direct signal from the source should be used in estimating the \gls{tdoa}. A common approach using e.g. GCC-PHAT is to find the strongest correlation peak and potentially putting some requirements on the strength of this peak. 

Because the \gls{impd}s are circular variables, the estimation of \gls{tdoa} requires solving a circular-linear fit \cite{Fisher}. For a probabilistic interpretation of the regression problem using wrapped IPDs, we refer to \cite{Traa2014}. However, since we are only considering frequencies below $f_u$, hereby avoiding phase ambiguity, an ordinary linear fit can be used as an approximation. In a commonly used least mean square fit, it is assumed that all data is pulled from a common distribution. However, for each mean \gls{impd}, a mapped mean resultant length is estimated, corresponding to a reliability measure of the mean \gls{impd}. Due to the aforementioned small inter-microphone spacing in the hearing aid setup, we employ the mapped mean resultant length in \eqref{eq:unwrapR} instead of the mean resultant length. 

Assuming for simplicity that the \gls{impd} follows a wrapped normal distribution, the variance ($\sigma_{a b}^2$) is given by \cite{Fisher},
\be
\sigma_{a b}^2(k,l) = - 2\log (\tilde{R}_{a b}(k,l)).
\label{eq:sigma}
\ee
For small variances a wrapped normal distribution is well approximated by a normal distribution. However, for small sample sizes, the low mean resultant length values are overestimated, corresponding to an underestimation of the variance, which leads to over emphasizing uncertain data points in the fit. As one way to circumvent this problem, we emprically found that using circular dispersion \cite{Fisher}, defined as 
\be
\delta_{a b}(k,l) = \frac{1-\tilde{R}^4_{a b}(k,l)}{2 \tilde{R}^2_{a b}(k,l)}
\label{eq:dispersion}
\ee
for a wrapped normal distribution, deemphasizes the uncertain data points. The reason for this is that $\delta_{a b}$ penalizes low $\tilde{R}$ values more than when using \eqref{eq:sigma}, while providing practically the same results for higher $\tilde{R}$ values. Considering that each data point has a known variance given by the circular dispersion and approximating the wrapped normal distribution with the normal distribution, the best least mean square fitted $\tau_{a b}$ takes the form 
\be
\tau_{a b}(l) = \frac{1}{2 \pi}\frac{\sum\limits_{k=1}^{K^{\prime}} \frac{\hat{\theta}_{a b}(k,l) f_k}{\delta_{a b}(k,l)} }{\sum\limits_{k=1}^{K^{\prime}} \frac{f_k^2}{\delta_{a b}(k,l)} },
\label{tau}
\ee
where $k$ is the frequency bin index, $\hat{\theta}_{a b}$ is the estimated mean \gls{impd} from \eqref{eq:mpd} and the summation higher limit $K^{\prime}<K$ denotes the number of frequency bins over which the fit is performed. The actual frequency is $f_k = f_s k / (2K)$. The variance of the estimated \gls{tdoa} can, by approximating $\delta_{a b}$ as a deterministic variable, be written as 
\be
\operatorname{var} \left( \tau_{a b} (l)\right) = \frac{1}{4 \pi^2} \frac{1}{\sum\limits_{k=1}^{K^{\prime}} \frac{f_k^2}{\delta_{a b}(k,l)}}.
\label{var}
\ee
This expression contains a number of simplifications and it should only be considered as an approximation. However, using \eqref{var} allows for a computationally simple closed form approximation of the variance of the estimated \gls{tdoa}, which can be utilized throughout the further stages to associate data based on their variance. 
%The approximation of considering $\sigma^2_{a b}$ being non stochastic is renders Eq. (\eqref{var}) approximative but allow for a simple traceable estimate of the variance of the \gls{tdoa}. 
%The variance ($\sigma^2$) used in Eq. (\ref{tau}) and (\ref{var}) are the variance of normal distributions are not directly equivalent to the circular variance (introduced in Sec. \ref{sec:phase}). The circular statistical equivalent is the variance of at wrapped normal distribution. Assuming that the phase for each time-frequency bin is wrapped normal distributed with mean phase $\hat{\theta}$ the variance $\rho^2$ is given fully by the mean resultant length \cite{Fisher}
%\be
%dsadas
%\ee 
%however since the correct variance to use is the circular dispersion.... 
%I should just use the ordinary circular standard deviation ($\rho^2$) and include a hyposistest for uniformity to take care of frequencies which are uniform (difuse dominated!!!!)... also try to run the code and see how it works!!!!!!! this is instead of using the circular dispersion
%
%Note that the non periodic linear fit described above is only valid for frequencies below phase ambiguity limit (add numbers), however for all microphone combination in the dummy head setup these limits are either equally or less conservative then the frequency range for which \gls{ff} can be assumed. 
\subsection{Monaural and binaural information integration}\label{sec:fusion} 
From the estimated \gls{tdoa} and its variance, a local \gls{doa} can be estimated for each microphone pair along with its variance. In the proposed method only azimuth \gls{doa} is considered and the look direction of the hearing aid user is defined as zero. Three microphone pairs are required in \gls{cimpl}: the two (left and right) monaural combinations ($M \in \{L,R\}$) and a binaural ($B$) pair. Additional binaural pairs can be included to improve the accuracy. Assuming far and free field and that the monaural arrays point in the look direction, the local \gls{doa}s can be estimated from the monaural \gls{tdoa}s as follows,
\be
\phi_M = {\arccos} \left( \frac{c}{d_M} \tau_M \right), 
\label{eq:mapdoaM}
\ee
where $d_M$ is the inter-microphone spacing between the two microphones on one hearing aid (monaural). Note that, even though the calculations take place at each frame $l$ (i.e., $\phi_M \equiv \phi_M(l)$), here and in the rest of the paper we drop the time index for conciseness. Using the Taylor expansion of \eqref{eq:mapdoaM} around $\phi_M=90^{\circ}$, the variance of the estimated monaural \gls{doa}s can be approximated from the variance of the \gls{tdoa}s as
\be
\operatorname{var} \left( \phi_M \right) \approx \left( \frac{c}{d_M} \right)^2 \operatorname{var} \left( \tau_M \right),
\label{eq:varM}
\ee
where the $\operatorname{var} \left( \tau_M \right)$ is estimated using \eqref{var}. 

For the binaural microphone pair, we assume far field and an ellipsoidal head model \cite{Duda1999}. From this, the binaural \gls{doa} is well approximated by
\be
\phi_B \approx \left( \frac{c}{d_B} \tau_B \right),
\label{eq:mapdoaB}
\ee
where $d_B$ is the inter-microphone spacing between the two hearing aids on the head and the look direction is perpendicular to the rotation axis of the binaural microphone pair. The variance of the estimated binaural \gls{doa} can be written as
\be
\operatorname{var}\left( \phi_B \right) = \left( \frac{c}{d_B} \right)^2 \operatorname{var} \left( \tau_B \right).
\label{eq:varB}
\ee

The estimated \gls{doa}s are circular variables and their estimated variances are transformed to mean resultant lengths using \eqref{eq:sigma}, where each \gls{doa} is assumed to follow a wrapped normal distribution. We denote $R_M$ ($M \in \{L,R\}$) and $R_B$ as the monaural and the binaural mean resultant lengths associated with the angle of arrivals, respectively.

The monaural \gls{doa} estimates  for the left and the right pairs are defined in the interval $[0,\pi]$ due to the rotational symmetry around the line connecting the microphones. Correspondingly, the binaural \gls{doa} is defined within $[-\pi /2,\pi /2]$. In order to combine the information from the monaural pairs and the binaural pair, a common support must be established. This is accomplished by mapping all azimuth estimates onto the full circle ($\varphi \in [-\pi,\pi[$). The choice of the monaural mean resultant length depends on which hearing aid is closer to the source. Using the binaural pair, we determine whether a given source is to the left ($\phi_B \ge 0$) or the right ($\phi_B < 0$). Based on this, if the source is located on the left, the left monaural microphone pair is chosen ($\varphi_M = \phi_{L}$), and similarly on the right side ($\varphi_M = -\phi_{R}$). Due to the head shadow effect, the monaural microphone pair closer to the source yields a more reliable estimate. From the chosen monaural pair it can be determined if a potential source is in front of ($|\varphi_M| \le \pi/2 $) or behind ($|\varphi_M| > \pi/2 $) the hearing aid user. When a source is in the front, then $\varphi_B = \phi_B$. If the source is determined to be to the right and behind the wearer, then $\varphi_B = -\pi - \phi_B$, and if it is behind and to the left, then $\varphi_B = \pi - \phi_B$. The mean resultant lengths are invariant under translations and are converted directly. 

We have a monaural and a binaural azimuth estimate of the full-circle \gls{doa} with their mean resultant lengths. From this, a statistical test is performed to assess the null hypothesis that the two estimates have a common mean \cite{Fisher}. The modified test statistic that we employ is
\begin{eqnarray}
Y & = & 2 \left( \left(\frac{w_M}{\delta_M} +\frac{w_B}{\delta_B} \right) - \sqrt{C^2 + S^2} \right),
\end{eqnarray}
where $C$ and $S$ are given by
\begin{eqnarray}
C & = & \frac{w_M}{\delta_M} \cos(\varphi_M) + \frac{w_B}{\delta_B} \cos(\varphi_B),\\ \nonumber
S & = & \frac{w_M}{\delta_M} \sin(\varphi_M) + \frac{w_B}{\delta_B} \sin(\varphi_B). \nonumber
\end{eqnarray}
Here, $\delta$ is the circular dispersion known from \eqref{eq:dispersion}, $w_M = {\sin}^2(\varphi_M)$ and $w_B = {\cos}^2(\varphi_B)$ are weighting factors for the monaural and binaural estimates, respectively, and $Y$ is the test statistic to be compared with the upper 100(1-$\alpha$)\% point of the $\chi^2_1$ distribution, with $\alpha$ as the significance level. The weighting factors are used to effectively reduce the reliability of the estimates to compensate for the approximations made in \eqref{eq:varM} and \eqref{eq:varB}. %Note that, in the above tests, the sample sizes of the monaural and the binaural pairs are assumed to be equal to one i.e. that the circular dispersion and the circular standard error are the same. 
If the null hypothesis is accepted with $\alpha = 0.1$, a common mean direction $\hat{\varphi}$ of the two estimates is calculated as \cite{Fisher}
\begin{eqnarray}
\hat{\varphi} & = & \angle \{ w_1 R_M e^{i \varphi_M} + w_2 R_B e^{i \varphi_B} \},
\label{eq:commonmean}
\end{eqnarray}
with
\begin{eqnarray}
w_1 & = & \frac{w_M/\left(R_M \delta_M\right)}{w_M/\left(R_M \delta_M\right)+w_B/\left(R_B \delta_B\right)} , \nonumber \\
w_2 & = & \frac{w_B/\left(R_B \delta_B\right)}{w_M/\left(R_M \delta_M\right)+w_B/\left(R_B \delta_B\right)} . \nonumber \\
\end{eqnarray}
Similarly, the circular dispersion of the common mean direction is
\begin{eqnarray}
\delta & = & 2 \frac{w_1^2 R_M^2 \delta_M + w_2^2 R_B^2 \delta_B}{\left(w_1 R_M + w_2 R_B \right)^2}.
\label{eq:commondispersion}
\end{eqnarray}
Subsequently, the mean resultant length of the common mean can be calculated by solving \eqref{eq:dispersion} for $R$ using the circular dispersion obtained by \eqref{eq:commondispersion} yielding
\begin{eqnarray}
R = \frac{1}{\sqrt{\delta+\sqrt{1+\delta^2}}}.
\label{eq:commonR}
\end{eqnarray}

If the null hypothesis is rejected, the \gls{doa} and its mean resultant length are chosen from the estimate with the lowest circular dispersion, i.e., either the monaural or the binaural. 

From the above development, the information provided from the monaural and the binaural \gls{tdoa}s and their variance are combined to make a unified full-circle \gls{doa} $\hat{\varphi}$ estimate in \eqref{eq:commonmean} with an accompanying circular dispersion $\delta$ in \eqref{eq:commondispersion} and the mean resultant length $R$ in \eqref{eq:commonR}. 
\subsection{Source tracking}\label{sec:tracker} 
The azimuth estimation at the output from the previous stage is very noisy, but at the same time it is accompanied by an instantaneous indication of reliability in the form of the mean resultant length $R$ \eqref{eq:commonR} or the circular dispersion \eqref{eq:commondispersion}. We include an angle-only wrapped Kalman filter \cite{traa2013wrapped} to obtain a smoother estimate. Differently from the original method described in \cite{traa2013wrapped}, which assumes a fixed and known variance denoted by $\sigma_w^2$ for the innovation term, we update this quantity at each frame using the circular dispersion as an approximation, i.e. $\sigma^2_{w_t} \approx \delta$. By using circular dispersion provided in \eqref{eq:commonR} instead of variance, low $R$ values map onto higher $\sigma^2_w$ values.
%, and the wrapped normal distribution is always fairly approximated through the sum of three normal distributions \cite{traa2013wrapped}.

\section{Evaluation}\label{sec:results}
%As noted in the previous section, due functional form of the \gls{pm} direct evaluation of performance and robustness is difficult. 
The LOCATA challenge development dataset \cite{LOCATA2018a} was used to assess the performance of \gls{cimpl}. More specifically, the hearing aid recordings in the presence of a single static source (task 1), a single moving source (task 3) and a single moving source with a moving listener (task 5) were considered.

\begin{figure}[h]%
\centering
\includegraphics[width=8.5cm]{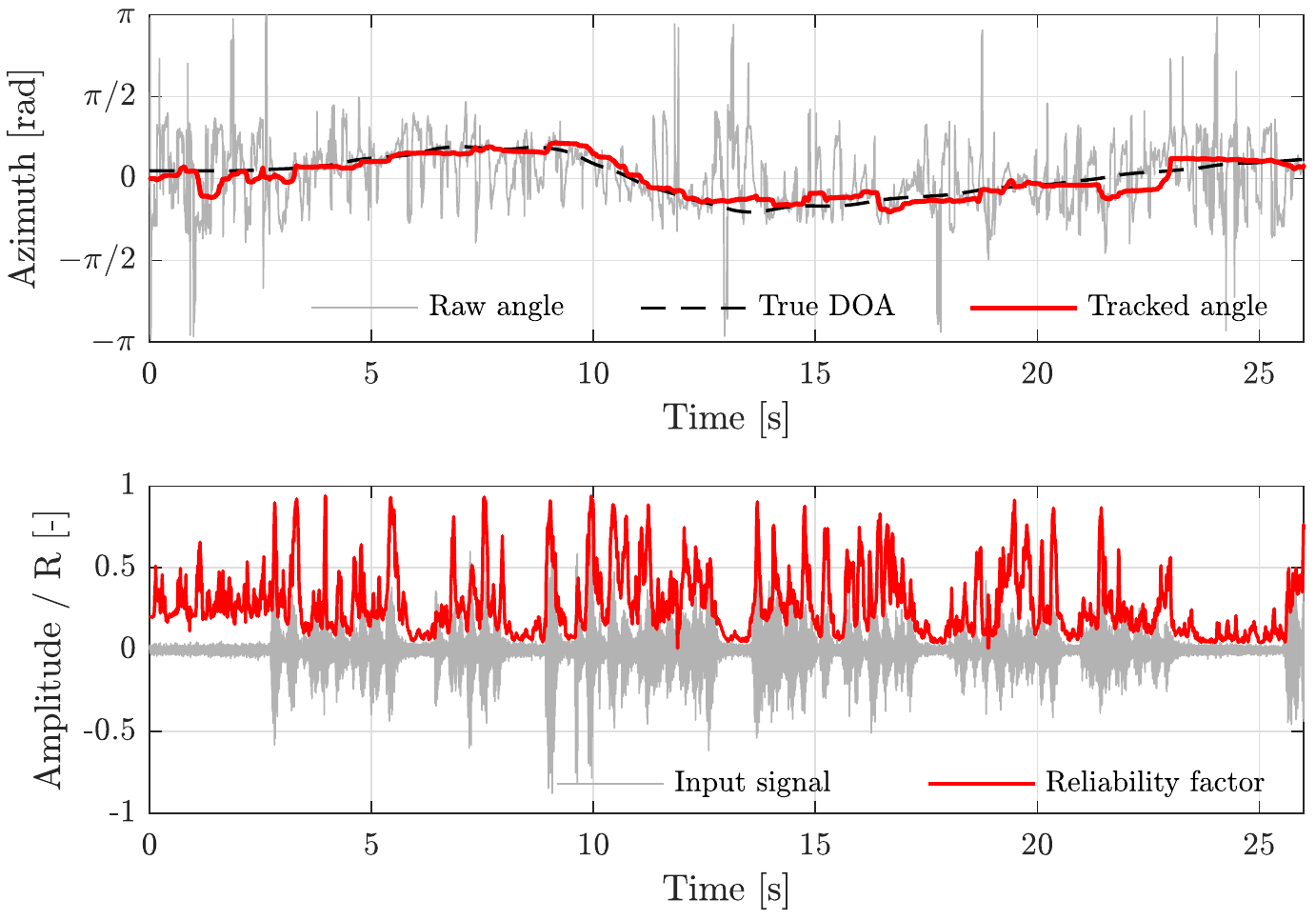}%
% \vspace{-0.3cm}
\caption{[Top] Azimuth tracking of a single moving source with \gls{cimpl} (red) and ground truth (dashed), together with raw angle estimates before the wrapped Kalman filter (gray). [Bottom] Raw audio signal (gray) and the reliability factor (red) used as input to the wrapped Kalman filter.}%
\label{fig:results_task3}%
\end{figure}

The standard deviation of the process noise in the wrapped Kalman filter was set to 1$^\circ$. Figure~\ref{fig:results_task3} illustrates the behavior of the algorithm for a recording of a single moving source. Notice that the raw azimuth estimates, shown in gray on the top panel, were very noisy. In contrast, the tracked angles, shown in red on the top panel, are smoother and more accurate thanks to the use of a wrapped Kalman filter. The input measurement variance to the wrapped Kalman filter was updated at each frame with the dispersion $\delta$, related to the reliability factor of the estimates, shown in red on the bottom panel, shown in Figure~\ref{fig:results_task3}. % through \eqref{eq:commonR}.

The mean absolute deviation from the ground truth (with standard deviation shown in parentheses), averaged across all data segments where speech was active, was 5.9$^\circ$ (10.4$^\circ$) for task 1, 8.2$^\circ$ (8.2$^\circ$) for task 3, and 18.7$^\circ$ (23.5$^\circ$) for task 5. 
\begin{figure}[h]%
\centering
\includegraphics[width=8.5cm]{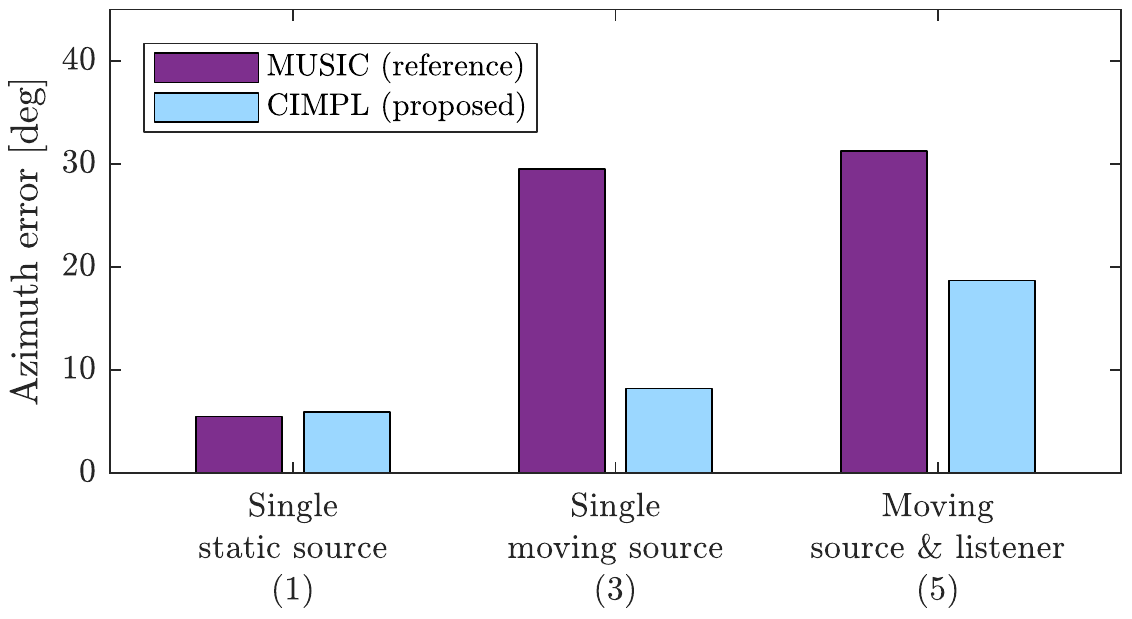}%
% \vspace{-0.3cm}
\caption{Azimuth accuracy for Tasks 1, 3 and 5 for the hearing aid recordings of the LOCATA challenge development dataset \cite{LOCATA2018a}.}%
\label{fig:results_tracking}%
\end{figure}

As shown in Figure~\ref{fig:results_tracking}, the performance of \gls{cimpl} in task 1 is comparable to that provided by the tracked MUSIC algorithm provided by LOCATA Challenge \cite{LOCATA2018a} as the benchmark, and better in tasks 3 and 5. Moreover, \gls{cimpl} runs in 1.3\% of the CPU time required by the tracked MUSIC algorithm \cite{LOCATA2018a} provided in the LOCATA challenge.

\section{Concluding remarks}\label{sec:discuss}
% conclusion remarks 
In this paper we proposed a new \gls{doa} estimator targeted for tracking a single source with a binaural hearing aid setup. By estimating the angle via circular statistics, the mean resultant length is obtained which acts as a reliability index. The mean resultant length is then carried throughout all the processing steps and is used at the tracker to improve the accuracy of the tracked angle. 
% \textbf{The method consists of the following steps: \gls{impd} estimation based on circular statistics, \gls{tdoa} estimation for monaural and binaural microphone pairs by fitting a straight line to the \gls{impd}s in the frequency domain, mapping \gls{tdoa}s into local \gls{doa}s, integration of local estimates into a single global \gls{doa} estimate and a wrapped Kalman filter as source tracker.}

Performance evaluation of the proposed method on the hearing aid recordings provided in the development dataset of the LOCATA challenge~\cite{LOCATA2018a} revealed an improved accuracy of the tracked angle of a single moving source compared to the benchmark method (tracked MUSIC algorithm) provided by the organizers, while running approximately 75 times faster. The low computational complexity of our algorithm makes it a favorable choice for hearing aid application. The estimated angle may be used at further stages of potential hearing aid processing, such as informed beamforming or scene classification. %The patents covering the CIMPL method and its applications for hearing aid use has recently been filed \cite{Wpatent1,Wpatent2}\textcolor{red}{@@TODO: TWO PATENTS CITATION}.

\balance

\bibliographystyle{IEEEtran}
\bibliography{refs}

\end{document}